\documentclass[11pt]{iopart}
\usepackage{epsfig}
\usepackage{graphics}
\usepackage{graphicx}
\usepackage[square,numbers,sort&compress]{natbib}
\usepackage{iopams}
\usepackage{bm}
\usepackage{color}
\usepackage[utf8]{inputenc}
\begin{document}

\title[van der Waals interactions in first principles topological insulator calculations]{Importance of van der Waals interactions for ab initio studies of topological insulators}

\author{K Shirali$^1$, W A Shelton$^{2, 3}$, and I Vekhter$^1$}

\address{$^1$ Department of Physics and Astronomy, Louisiana State University, Baton Rouge, LA 70803-4001}
\address{$^2$ Cain Department of Chemical Engineering, Louisiana State University, Baton Rouge, LA 70803-4001}
\address{$^3$ Center for Computation and Technology, Louisiana State University, Baton Rouge, LA 70803-4001}
\eads{\mailto{kshira1@lsu.edu}, \mailto{wshelton@lsu.edu}, \mailto{vekhter@lsu.edu}}


\begin{abstract}
We investigate the lattice and electronic structures of the bulk and surface of the prototypical layered topological insulators Bi$_2$Se$_3$ and Bi$_2$Te$_3$ using ab initio density functional methods, and systematically compare the results of different methods of including van der Waals (vdW) interactions. We show that the methods utilizing semi-empirical energy corrections yield accurate descriptions of these materials, with the most precise results obtained by properly accounting for the long-range tail of the vdW interactions. The bulk lattice constants, distances between quintuple layers and the Dirac velocity of the topological surface states (TSS) are all in excellent agreement with experiment. In Bi$_2$Te$_3$, hexagonal warping of the energy dispersion leads to complex spin textures of the TSS at moderate energies, while in Bi$_2$Se$_3$ these states remain almost perfectly helical away from the Dirac point, showing appreciable signs of hexagonal warping at much higher energies, above the minimum of the bulk conduction band. Our results establish a framework for unified and systematic self-consistent first principles calculations of topological insulators in bulk, slab and interface geometries, and provides the necessary first step towards ab initio modeling of topological heterostructures.
\end{abstract}

\noindent{\it Keywords\/}: topological insulators, density functional theory, topological surface states

\submitto{\JPCM}

\maketitle

\section{Introduction} Most proposed applications of topological insulators (TIs) involve fabricating heterostructures combining these materials with other topologically trivial compounds, and creatively controlling the spin-momentum locked symmetry-protected states at the interface between the two. These interface states are often assumed to be helical and isotropically dispersing, in resemblance to their counterparts at surfaces~\cite{Kane_RMP_2010,Zhang_RMP_2011}. Theoretical models, however, show that symmetry breaking at the interface leads to substantial modifications of the properties of topological states~\cite{Zhang2012,Mahmoud2017}. In real systems such symmetry breaking interface potentials may originate from strain and stress due to lattice mismatch, broken bonds, buckling, and surface reconstruction. Experiments indicate that they do indeed modify, sometimes drastically, the expected behavior of the topological states~\cite{QZhang2012,Zeljkovic2015,Richardella2015}. To gain a detailed understanding of the origin, form, and magnitude of these potentials, we first need to select and implement methods which treat the structural and electronic properties of both the bulk and surface states on equal footing. This step would lead to development of a comprehensive picture that serves as a reference and starting point for the interface problem, and enables predictive use of ab initio methods for design of topological interfaces. It may therefore seem surprising that there has been no clear agreement on how precisely to take this critical step.

The main structural unit of prototypical TIs of the Bi$_2$X$_3$ (X=Se,Te) family is a ``quintuple layer'' (QL) X$^\prime -$Bi$-$X$-$Bi$-$X$^\prime$, where each atom represents a layer, and the respective pairs of X$^\prime$ and Bi positions are related through inversion symmetry. Since each QL effectively has a ``closed shell''~\cite{Lind2005}, the coupling between QLs is believed to be largely due to van der Waals  forces. However, the electronic band dispersion in the direction normal to the layers is comparable to that in the plane, see below, indicating that vdW interactions can be treated as a correction in the ab-initio calculations of the properties of the TIs.

An important question is which of the many available forms of the vdW corrections satisfactorily describes both structural and electronic properties at the same time, and we comprehensively address it here within the framework of the density functional theory (DFT)-based ab-initio methods. While vdW interactions in layered TIs have been investigated, a complete picture has not been presented, and we fill this gap.

Many previous DFT studies of Bi$_2$X$_3$ in bulk and at surfaces~\cite{Chang2015} used experimentally determined lattice constants~\cite{Yazyev2010,Yazyev2012,Crowley:2015,Sun2017,Michiardi2014,Youn2001,Chen2011}, fixing the volume of the unit cell, and finding the relaxed atomic positions within this cell under the constraint of maintaining the crystal symmetry. The results reproduce salient features of the electronic spectra in bulk and at surfaces, although there is some debate about whether many body corrections are necessary to obtain both the correct magnitude of the gap and its character (direct vs indirect)~\cite{Nechaev2013,Aguilera2013,Michiardi2014}. In the same spirit, existing studies of interfaces
either consider lattice matched cases~\cite{HLee:2017} or make a priori assumptions about structural changes at the boundary, such as fixing the inter-QL distance to its experimental value~\cite{Zhang2016}.

The danger inherent to such approaches is that the standard DFT is a ground state theory. Applying it to a fixed set of parameters without allowing their values to attain their ground state values within the same theory may, in general, give the results for, e.g. electronic structure, that are not characteristic of the ground state of the material being studied.

There have been two principal reasons for continuing to make these assumptions. First, while it is known that full geometry relaxation of the bulk (when the unit cell volume is allowed to change) using standard exchange-correlation functionals such as the Local Density Approximation (LDA) or the Generalized Gradient Approximation (GGA) yields lattice parameters significantly different from those determined experimentally, which implies that the strain field is not correctly determined using these methods~\cite{Lind2005}, using the experimental values allowed a quick (and perhaps somewhat fortuitous) access to the qualitative salient features.  Second, since the electronic structure of bulk Bi$_2$X$_3$ only weakly varies (on the scale of the bandwidth) when vdW and similar corrections to the GGA and LDA are included, and since the existence of the topological Dirac state at the surface is protected by symmetry, there seemed to be  little incentive to include them for basic analysis. At interfaces in prototype devices, however, the situation is different: strain leads to surface reconstruction and symmetry breaking, and changes the behavior of the topological states, so we need to be able to optimize the structure and investigate the dispersion and spin properties of the topological interface states within the same methodology. This is why we perform here a systematic investigation of the inclusion of vdW corrections in ab initio calculations of Bi$_2$X$_3$.

Other studies previously considered vdW interactions in first principles calculations of the bulk~\cite{Luo2012,Cheng2014,Bjorkman2012,Reid2020} and surfaces~\cite{Reid2020,YZhang_vdW:2015,Govaerts2014,Liu2013}.
One of them~\cite{Luo2012} did not treat the lattice and electronic structure in the same framework: the authors first performed structural optimization including vdW but omitting the spin-orbit interaction (SOI), well known to be critical for band inversion in topological insulators; they then computed the electronic band structure with SOI (employing GGA), not changing the lattice. Other publications~\cite{YZhang_vdW:2015,Govaerts2014,Liu2013,Reid2020} addressed the properties of the surface states considering, as is common in the ab-initio work, periodic structure consisting of a slab of a TI surrounded by vacuum. However, they used slab geometries with small vacuum thickness, and our results below imply that in those cases slabs interacted with each other, suggesting that the results were not applicable to a free surface. Indeed, some of these works obtained lattice constants and electronic dispersion which differ from both experiment and our results presented below, and are not compared to bulk results. The effects of strain on slabs of Bi$_2$Se$_3$, Bi$_2$Te$_3$, Sb$_2$Te$_3$ and Sb$_2$Se$_3$ have been investigated, including vdW interactions in the first principles calculations~\cite{Aramberri2017}. However, the authors of this investigation~\cite{Aramberri2017} only considered the DFT+D2 method, and did not compare with other techniques. 
The results of a study~\cite{Bjorkman2012} on a large number of layered materials, which included the TIs that are the focus of our investigation, considered different vdW methods in the bulk, but did not analyze surface properties, and hence did not consider the slab geometry. As a result, the authors could not make a clear connection with surface and topological properties.
We are aware of only one previous investigation~\cite{Cao2018}, which considered several different implementations of vdW interactions in density functional theory, compared the outcomes, and argued for the most appropriate method for layered TIs. Our conclusions about the optimal methodology, as is clear from the remainder of the paper, differ from theirs, and we provide a physical picture that supports our findings.

Meta-GGA functionals promise a better description of several classes of materials within the DFT approach, and a recent paper~\cite{Reid2020} reported the results for SCAN+vdW in comparison with GGA/LDA with and without D3 corrections for the 3D TIs. Their results for SCAN differ from ours, as discussed below: most dramatically, we find that this method yields a metallic ground state for the 3D TIs that we considered.

Here we carry out a comprehensive comparison of different implementations of vdW interactions in bulk and surface DFT calculations of Bi$_2$Se$_3$ and Bi$_2$Te$_3$. We first show that the results of the LDA and GGA calculations are unable to fully describe the physics governing the structural and electronic properties of layered TIs. We then consider two classes of vdW methods. The first class accounts for the van der Waals interactions by adding a semi-empirical correction to the energy calculated using an exchange correlation functional such as GGA or LDA, and includes DFT+D2~\cite{Grimme2006}, DFT+D3~\cite{Grimme2010}, Tkatchenko-Scheffler~\cite{Tkatchenko2009,Bucko2013} (TS), and Tkatchenko-Scheffler with Many-body Dispersion~\cite{Tkatchenko2012,Ambrosetti2014} (TS-MBD). The second class of methods uses functionals which contain a nonlocal long-range vdW correlation, where the correlation depends on the electron density and its gradient, and includes, in our analysis, SCAN-rVV10~\cite{Sun2015,Peng2016} and Langreth-Lundqvist vdW-DF2~\cite{Lee2010}.

We do not use here hybrid functionals~\cite{Crowley:2015} that take into account a fraction of the exact Hartree-Fock exchange energy, nor do we emply the techniques that compute the electron self-energy with the renormalized Coulomb potential (GW)~\cite{Yazyev2012,Nechaev2013}, which aim to improve the accuracy of these approximations. Both the (in principle very precise) GW method and its non-self-consistent counterpart, G$_0$W$_0$, are computationally expensive, rendering them impractical for surface and interface calculations~\cite{Crowley:2015}. Our goal is to set up a comprehensive framework which treats structural and electronic properties on equal footing, which is necessary for first principles calculations of interfaces.

To evaluate the accuracy of different methods we compare the values for the bulk parameters, including, in addition to the lattice constants,  the inter-QL distance $d_{int}$, as well as the band gap and the electronic density of states. We also used the value of the Dirac velocity for the surface states as a test of the quality of our approaches for the slab calculations.

We find that the semi-empirical methods are consistently more
accurate than the the vdW functionals in yielding the lattice and electronic structures, as well as the correct {and stable} properties of the surface states for both Bi$_2$Se$_3$ and Bi$_2$Te$_3$. The unit cell volume and the electronic density of states obtained using vdW-DF2 functional differ significantly both from the experimental values and from the results obtained using semi-empirical methods. The SCAN functional and its variations predict Bi$_2$X$_3$ to be a metal.

The rest of the paper is organized as follows. Section 2 specifies the computational details and methodology for the bulk and slab calculations. Section 3 describes the results of the bulk calculations, and Section 4 builds on these results to provide a description of the topological surface states by doing the calculations in the slab geometry. 
Finally, in Section 5 we place our results in the context of other work, and discuss their implications of our results on 
for first principles calculations of interfaces of topological insulators. 

\begin{figure}
    \centering
    \includegraphics[width=\columnwidth]{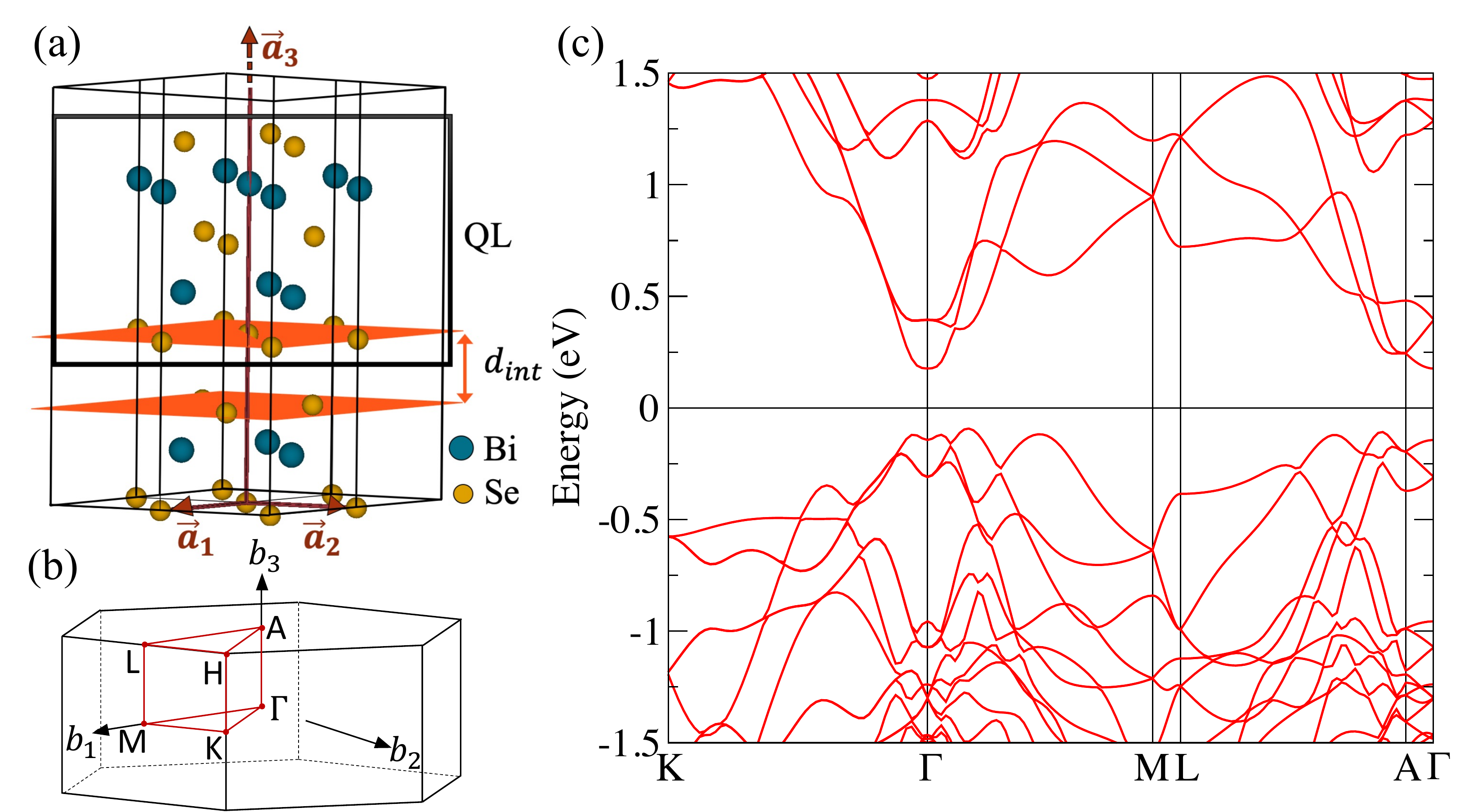}
    \noindent\caption{\label{fig:Bulk} Bulk structure and electronic properties of Bi$_2$Se$_3$. (a) Crystal structure with hexagonal unit cell, including the inter-QL distance, $d_{int}$. 
    We show the in-plane lattice vectors $\vec{\bm a}_{1,2}$, but the dashed arrow for lattice vector $\vec{\bm a}_3$ indicates that it extends beyond the range shown, as the unit cell in this representation contains 3 QLs. (b) Hexagonal Brillouin zone. (c) Electronic band structure calculated using GGA with van der Waals interactions (DFT+D2 method).} 
\end{figure}

\section{Computational Details} We used hexagonal unit cells for Bi$_2$Se$_3$ and Bi$_2$Te$_3$ (see figure~\ref{fig:Bulk}(a)) which are convenient for the description of the surface states.
All calculations below were carried out using the Vienna Ab initio Simulation Package~\cite{Kresse:1993,Kresse:1994,Kresse:1996a,Kresse:1996b} (VASP), version 5.4.4. Crystallographic information is taken from experimental data~\cite{Nakajima1963} retrieved from Crystallography Open Database~\cite{Merkys2016, Merkys2015,Merkys2012,Grazulis2009,Downs2003}. 
In our calculations we use data on the Bi$_2$Se$_3$ and Bi$_2$Te$_3$ crystal structure of Nakajima~\cite{Nakajima1963}, rather than earlier data~\cite{semiletov1955electron,wyckoff1964structures}, based on a recent analysis~\cite{Luo2012}.
We used Project Augmented Wave (PAW) potentials~\cite{Blochl:1994,Joubert:1999} for Bi ($5d^{10} 6s^{2} 6p^{3}$) and Se ($4s^{2} 4p^{4}$), for a total of 48 electrons, and a plane-wave basis. Convergence tests revealed that a $\Gamma$-centered k-point grid of 11x11x11 k-points and an energy cut-off of 450 eV for the plane wave basis are sufficient for high accuracy results. We performed fully relativistic calculations which include spin-orbit interaction (SOI). The convergence threshold for energy is taken to be $10^{-5}$ eV. Band structures are plotted with data processed using vaspkit~\cite{vaspkit}.  We used GGA--PBE~\cite{PerdewBurkeErnzerhof:1996,PerdewBurkeErnzerhof:1997}, LDA~\cite{Perdew:1981}, LDA+U and GGA+vdW, with  full relaxation; for the last category, we compared the results of several methods including van der Waals interactions: DFT--D2~\cite{Grimme2006}, DFT--D3~\cite{Grimme2010}, DFT--TS~\cite{Tkatchenko2009,Bucko2013} (Tkatchenko-Scheffler), DFT--TS-MBD~\cite{Tkatchenko2012,Ambrosetti2014} (Tkatchenko-Scheffler with Many-Body Dispersion). The vdW functionals studied consisted of the SCAN~\cite{Sun2015}, SCAN--rVV10~\cite{Peng2016} and vdW-DF2~\cite{Lee2010} methods.  

Semi-empirical methods add to the total energy corrections proportional to $r_{ij}^{-6}$ (DFT+D2, DFT+TS) with additional terms varying as $r_{ij}^{-8}$ (DFT+D3) for each pair of atoms $i,j$ which are separated by less than a cutoff distance. To reduce the contribution from  pairs of atoms that are bonded covalently, these methods employ a short-distance damping function.
The method DFT+TS uses the same energy correction as DFT+D2, modifying the damping function and dispersion coefficients in the energy correction to reflect the local chemical environment around each atom. The method DFT+TS-MBD (MBD@rsSCS) uses a random phase approximation-like expression for the self-consistent screening, treating each atom as a fluctuating dipole, to arrive at the van der Waals energy correction.

The van der Waals functionals considered in this study are the SCAN, SCAN--rVV10, and the vdW-DF2 (Langreth - Lundqvist) functionals. SCAN is a type of 
semi-local exchange-correlation functional which includes the intermediate-range vdW interactions, while rVV10 is a nonlocal functional which accounts for the long-range vdW interactions by including a long range energy correlation term which depends on the electron density and its gradient. The vdW-DF2 method uses the exchange functional PW86 (which does not incorporate a van der Waals correction), and adds a nonlocal functional which contains the long-range van der Waals correction.

For slab calculations, the surfaces were modelled using slabs of thickness 5-7 QL ($\sim$ 50-70 \AA). We find that slabs 5 QL thick are sufficient since the hybridization gap between topological states at different surfaces is small. With vdW interaction included we find that using a vacuum thickness roughly equal to twice that of the slab is sufficient to avoid interaction of the slab with periodic images of itself. Atoms in the outermost QLs of the slab are allowed to relax in all directions without restriction, while atoms in the ``bulk'' part of the slab are held fixed. Slab calculations were run using the methods DFT+D2, DFT+D3, DFT+TS and DFT+TS-MBD and compared based on features of the surface states such as the Dirac velocity and spin texture. A dipole correction along the $z$ direction was tested and found to not contribute significantly, which is consistent with the picture of closed shell QLs.

\begin{table}
\caption{\label{Bi2Se3_bulk_LDAGGA} Results for the structural optimization of Bi$_2$Se$_3$ without the van der Waals corrections, using GGA and LDA. For each parameter we show the percentage difference relative to the experimental value.}
\footnotesize
\begin{indented}
\item[]\begin{tabular}{@{}|l|l|l l|l l|}
 \br
  & Expt. & \multicolumn{2}{c|}{GGA}  & \multicolumn{2}{c|}{LDA} \\
  \hline
  & & & & & \\
 \hline
 $a$ ({\AA}) & 4.143~\cite{Nakajima1963} & 4.227 & +2.0\% & 4.086 & -1.4\% \\ 
 \hline
 $c$ ({\AA}) & 28.636~\cite{Nakajima1963} & 29.218 & +2.0\% & 28.244 & -1.4\% \\
 \hline
 $d_{int}$ ({\AA}) & 2.579~\cite{Nakajima1963} & 2.746 & +6.5\% & 2.378 & -7.8\%  \\
 \hline
 Gap $E_g$ (eV) & 0.3~\cite{Xia2009} & 0.1567 & -48\% & 0.4594 & +53\% \\
 \br
 \end{tabular}
 \end{indented}
 \end{table}

\begin{table}
\caption{\label{Bi2Se3_bulk_LDAU} Results for the structural optimization of Bi$_2$Se$_3$ without the van der Waals corrections, using LDA+U. For each parameter we show the percentage difference relative to the experimental value.}
\footnotesize
\begin{indented}
\item[]\begin{tabular}{@{}|l|l|l l|l l|l l|}
 \br
  & Expt. & \multicolumn{6}{c|}{LDA+U} \\
  \hline
  & & \multicolumn{2}{c|}{$U=3$} & \multicolumn{2}{c|}{$U=5$} & \multicolumn{2}{c|}{$U=7$} \\
 \hline
 $a$ ({\AA}) & 4.143~\cite{Nakajima1963} & 4.115 & -0.7\% & 4.144 & +0.024\% & 4.159 & +0.4\%  \\ 
 \hline
 $c$ ({\AA}) & 28.636~\cite{Nakajima1963} & 28.445 & -0.7\% & 28.646 & +0.035\% & 28.750 & +0.4\% \\
 \hline
 $d_{int}$ ({\AA}) & 2.579~\cite{Nakajima1963} & 2.429 & -5.8\% & 2.475 & -4.0\% & 2.514 & -2.5\%  \\
 \hline
 Gap $E_g$ (eV) & 0.3~\cite{Xia2009} & 0.2891 & -3.6\% & 0.2326 & -22\% & 0.1143 & -62\% \\
 \br
 \end{tabular}
 \end{indented}
 \end{table} 

\begin{table}
\caption{\label{Bi2Te3_bulk_LDAGGA} Results for the structural optimization of Bi$_2$Te$_3$ without the van der Waals corrections. For each parameter we show the percentage difference relative to the experimental value.}
\footnotesize
 \begin{indented}
\item[]\begin{tabular}{@{}|l|l|l l|l l|l l|l l|l l|}
 \br
  & Expt. & \multicolumn{2}{c|}{GGA}  & \multicolumn{2}{c|}{LDA} \\
 \hline 
 $a$ ({\AA}) & 4.386~\cite{Nakajima1963} & 4.476 & +2.1\% & 4.339 & -1.1\% \\ 
 \hline
 $c$ ({\AA}) & 30.497~\cite{Nakajima1963} & 31.124 & +2.1\% & 30.167 & -1.1\% \\
 \hline
 $d_{int}$ ({\AA}) & 2.613~\cite{Nakajima1963} & 2.783 & +6.5\% & 2.484 & -4.9\% \\
 \hline
 $E_g$ (eV) & 0.165~\cite{Chen2009} & 0.137 & -17\% & 0.103 & -38\% \\
 \br
 \end{tabular}
 \end{indented}
 \end{table}
 
\section{Bulk properties}
\subsection{LDA and GGA} 
We start with the results for the most commonly used LDA and GGA-PBE approximations that do not include vdW corrections. Our results for the structural optimization of Bi$_2$Se$_3$ (Bi$_2$Te$_3$) are shown in Table~\ref{Bi2Se3_bulk_LDAGGA} (Table~\ref{Bi2Te3_bulk_LDAGGA}). As often happens, LDA overbinds the electrons: for Bi$_2$Se$_3$, it leads to a contraction of the lattice constants compared to their experimental value. Our result for the volume change of the unit cell is in quantitative agreement with the value of 4\% found before~\cite{Lind2005}, albeit the values for the lattice constants differ due to different optimization procedures. In contrast, using GGA overestimates the unit cell volume by about 6\% in our calculation vs. a reported value of almost 10\%~\cite{Lind2005}. Another investigation~\cite{Luo2012} found the cell volume overestimated even more using PBE, but that is mostly due to a significant elongation of the $c$-axis lattice constant, perhaps related to the effectively two-dimensional k-point mesh (13x13x1) used in that work. 

Similarly, for Bi$_2$Te$_3$, LDA reduces the unit cell volume by 3\% compared to a value of 5\%~\cite{Luo2012}. GGA follows very similar trends to those in Bi$_2$Se$_3$. The 6.3\% increase in the unit cell volume is greater than a reported value of 4.9\%~\cite{Luo2012}. Another publication~\cite{Lawal2017} used GGA and found the out of plane lattice constant, \textit{c}, and inter-QL distance, $d_{int}$, which deviated $4.6\%$ and $18.7\%$ respectively from their experimental values. Our values for the corresponding parameters are smaller, see Table~\ref{Bi2Te3_bulk_LDAGGA}, possibly due to the different type of pseudopotential they used (fully relativistic norm-conserving).

Notably, using both GGA and LDA we find that the difference between the experimental values and those obtained from first principles is much greater for the distance between adjacent QLs (see figure~\ref{fig:Bulk}(a)), $d_{int}$ in Table~\ref{Bi2Se3_bulk_LDAGGA}, than for the lattice constants. This strongly suggests the need for inclusion of the van der Waals interactions between different quintuple layers. 

Both GGA and LDA yield sizeable deviations for the bulk energy gaps, as is common for small gap semiconductors with strong spin-orbit coupling~\cite{Perdew2009,Chan2010}. The band gap of 0.137 eV that we find for Bi$_2$Te$_3$ using GGA is slightly closer to the experimental value than a gap of 0.12 eV~\cite{Lawal2017}. Our band gap from an LDA calculation for Bi$_2$Te$_3$ is very close to the value of 0.106 eV reported for the LDA band gap~\cite{Youn2001}.

For completeness, and to clearly show that local Coulomb interactions are not the origin for the discrepancy, we performed LDA+U calculations on Bi$_2$Se$_3$ with the on-site repulsion on Bi orbitals, see Table~\ref{Bi2Se3_bulk_LDAU}. For $U=5$, the lattice constants $a$ and $c$ are very close to their experimental values, but the inter-QL distance still significantly deviates from that in experiment, and is only improved for $U=7$, while the lattice constants increase for that value. In contrast, the gap magnitude is closest to experiment for $U=3$. Therefore no single value of $U$ consistently improves the results. Moreover, increasing values of $U$ tend to collapse the energy gap, in contrast to the physical expectation that Coulomb repulsion localizes corresponding orbitals and pushes bands apart. 
The situation is even worse for GGA+U as we very quickly reach gap collapse and metallicity. We therefore conclude that the error in the inter-QL distance must be related to the long-range van der Waals part of the interaction.

 
 \begin{table}
\caption{\label{Bi2Se3_bulk_vdW_semiemp} Results for the structural optimization of Bi$_2$Se$_3$ using different semi-empirical van der Waals corrections. For each parameter we show the percentage difference relative to the experimental value.}
\footnotesize
\begin{indented}
\item[]\begin{tabular}{@{}|l|l|l|l|l|l|}
 \br
  & Expt. & DFT+D2 & DFT+D3 & DFT+TS & DFT+TS-MBD \\
 \hline
 $a$ ({\AA}) & 4.143~\cite{Nakajima1963} & 4.141 & 4.175 & 4.188 & 4.176 \\ 
 \hline
 Devn. $a$ & - & -0.05\% & +0.77\% & +1.1\% & +0.79\% \\ 
 \hline
 $c$ ({\AA}) & 28.636~\cite{Nakajima1963} & 28.624 & 28.858 & 28.948 & 28.866 \\
 \hline
 Devn. $c$ & - & -0.04\% & +0.77\% & +1.1\% & +0.80\% \\ 
 \hline
 $d_{int}$ ({\AA}) & 2.579\cite{Nakajima1963} & 2.547 & 2.587 & 2.623 & 2.584 \\
 \hline
 Devn. $d_{int}$ & - & -1.2\% & +0.31\% & +1.7\% & 0.19\% \\ 
 \hline
 Gap $E_g$ (eV) & 0.3~\cite{Xia2009} & 0.2638 & 0.2273 & 0.2125 & 0.2337 \\
 \hline
 Devn. $E_g$ & - & -12\% & -24\% & -29\% & -22\% \\ 
 \br
 \end{tabular}
 \end{indented}
 \end{table} 

\begin{table}
\caption{\label{Bi2Se3_bulk_vdW_functionals} Results for the structural optimization of Bi$_2$Se$_3$ using different van der Waals functionals. For each parameter we show the percentage difference relative to the experimental value.}
\footnotesize
\begin{indented}
\item[]\begin{tabular}{@{}|l|l|l|l|l|l|l|l|l|}
 \br
  & Expt. & SCAN & SCAN-rVV10 & vdW-DF2 \\
 \hline
 $a$ ({\AA}) & 4.143~\cite{Nakajima1963} & 4.182 & 4.177 & 4.489 \\ 
 \hline
 Devn. $a$ & - & +0.94\% & +0.82\% & +8.4\% \\ 
 \hline
 $c$ ({\AA}) & 28.636~\cite{Nakajima1963} & 28.907 & 28.872 & 31.029 \\
 \hline
 Devn. $c$ & - & +0.94\% & +0.82\% & +8.4\% \\ 
 \hline
 $d_{int}$ ({\AA}) & 2.579\cite{Nakajima1963} & 2.663 & 2.651 & 3.349 \\
 \hline
 Devn. $d_{int}$ & - & +3.3\% & +2.8\% & +30.0\% \\ 
 \hline
 Gap $E_g$ (eV) & 0.3~\cite{Xia2009} & Metal & Metal & 0.1749\\
 \hline
 Devn. $E_g$ & - & - & - & -42\%\\ 
 \br
 \end{tabular}
 \end{indented}
 \end{table} 

\begin{table}
\caption{\label{Bi2Te3_bulk_vdW} Results for the structural optimization of Bi$_2$Te$_3$ using different forms of van der Waals corrections. For each parameter we show the percentage difference relative to the experimental value.}
\footnotesize
 \begin{indented}
\item[]\begin{tabular}{@{}|l|l|l|l|l|l|l|}
 \br
  & Expt. & DFT+D2 & DFT+D3 & DFT+TS & DFT+TS-MBD & SCAN \\
 \hline
 $a$ ({\AA}) & 4.386~\cite{Nakajima1963} & 4.382 & 4.419 & 4.422 & 4.419 & 4.418 \\ 
 \hline
 Devn. $a$ & - & -0.09\% & +0.75\% & +0.82\% & +0.75\% & +0.73\% \\ 
 \hline
 $c$ ({\AA}) & 30.497~\cite{Nakajima1963} & 30.472 & 30.724 & 30.747 & 30.726 & 30.719 \\
 \hline
 Devn. $c$ & - & -0.08\% & +0.74\% & +0.82\% & +0.75\% & +0.73\% \\ 
 \hline
 $d_{int}$ ({\AA}) & 2.613~\cite{Nakajima1963} & 2.616 & 2.636 & 2.660 & 2.621 & 2.678 \\
 \hline
 Devn. $d_{int}$ & - & +0.11\% & +0.88\% & +1.8\% & +0.31\% & +2.5\% \\ 
 \hline
 Gap $E_g$ (eV) & 0.165~\cite{Chen2009} & 0.1608 & 0.1585 & 0.1572 & 0.1557 & Metal \\
 \hline
 Devn. $E_g$ & - & -2.5\% & -3.9\% & -4.7\% & -5.6\% & - \\
 \br
 \end{tabular}
 \end{indented}
 \end{table}
 
\subsection{Inclusion of van der Waals interactions} 
Including van der Waals corrections semi-empirically in the structural optimization leads to a dramatic improvement in the agreement with experiment for values of the lattice constants and the inter-QL spacing, see Table~\ref{Bi2Se3_bulk_vdW_semiemp}, Table~\ref{Bi2Se3_bulk_vdW_functionals} and Table~\ref{Bi2Te3_bulk_vdW}. 
The lattice constants deviate by less than about a percent from the experimental values, and the inter-QL spacing is also much closer to experiment. DFT+D2 produces particularly good results, where most deviations are below $0.1\%$, with the only exception being the inter-QL distance in Bi$_2$Se$_3$ which gives $1.2\%$ error, greater than the corresponding value of $-0.35\%$ found earlier~\cite{Luo2012}. 
The unit cell volume we find differs from experiment by $-0.14\%$ for Bi$_2$Se$_3$ and $-0.26\%$ for  Bi$_2$Te$_3$, compared to the respective reported values of $-0.07\%$ and $-1.14\%$~\cite{Luo2012} and $0.70\%$ and $0.82\%$~\cite{Aramberri2017} (obtained for a different crystal structure~\cite{wyckoff1964structures}). For Bi$_2$Te$_3$ a unit cell volume deviation of just $-0.07\%$ was found before~\cite{Cheng2014}, but the inter-QL distance in that calculation differed from experiment by $9.3\%$. In a calculation where the unit cell shape and volume were allowed to change~\cite{Lamuta2016}, the deviation in the unit cell volume was found to be $-1.8\%$. The significant difference with the latter result is likely due to the Monkhorst-Pack k-grid of 8x8x2 and scalar relativistic norm-conserving pseudopotentials they used. 

DFT+TS-MBD and DFT+D3 also produce good results for the structural parameters. The lattice constants and unit cell volume ($3.3\%$ deviation for Bi$_2$Se$_3$ and $2.5\%$ for  Bi$_2$Te$_3$) that we find using DFT+TS are close to the $2.9\%$ and $3.0\%$  values obtained in another study~\cite{Cao2018} for the same method.

Our results using vdW functionals are in much poorer agreement with the experimental values. The vdW-DF2 method significantly overestimates cell volume, bond lengths and underestimates the band gap, a behavior also reported for bulk Bi$_2$Te$_3$~\cite{Cheng2014}. That work argued in favor of using optB86b-vdW method, which yields a deviation in unit cell volume of $2.0\%$, and a deviation in inter-QL distance of $3.5\%$. Also for Bi$_2$Te$_3$, using vdW-DF resulted in a unit cell volume and inter-QL distance deviating $2.6\%$ and $1.5\%$ from experiment respectively~\cite{Lawal2017}. These deviations are an order of magnitude greater than what we find using DFT+D2.

We also carried out a calculation using a meta-GGA SCAN-rVV10 method. However, we found that for both materials it yields a metallic ground state. We checked that the SCAN method on its own also yields a metal as implemented in VASP, and confirmed this result for Bi$_2$Se$_3$ by checking that the same result is obtained in an all-electron calculation using Elk~\cite{elk}. Therefore we conclude that at least the current implementation of the SCAN functional is not suitable for describing layered topological insulators. This is in sharp contrast to a recently published study~\cite{Reid2020} in which a meta-GGA SCAN-based structural relaxation yields an insulator in that paper. Since that study~\cite{Reid2020} does not give full details of their calculations, and, for example, does not specify the $\bm k$-point grid for bulk calculations, we cannot make a detailed comparison of the methodologies leading to this discrepancy.

\begin{figure}
    \centering
    \includegraphics[scale=0.5,trim={1cm 1.5cm 1cm 1.5cm}, clip]{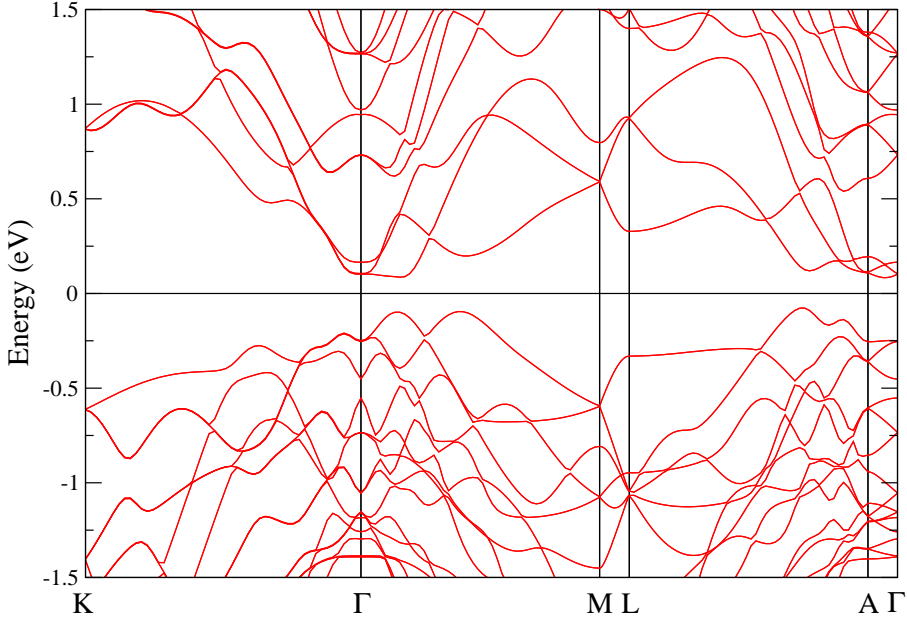}
    \caption{\label{fig:Bulk_Bi2Te3} Bulk electronic structure of Bi$_2$Te$_3$ calculated using GGA with DFT+D2 vdW method, see text for details.}
\end{figure}

Comparing the electronic band structures obtained by different methods provides a complementary check on their applicability. For Bi$_2$Se$_3$, the (indirect) band gap  obtained using DFT+D2 is much closer to the experimental value than that obtained using other functionals, see Table~\ref{Bi2Se3_bulk_vdW_semiemp}. The band structure shown in figure~\ref{fig:Bulk}(c)  exhibits, albeit not very strongly,  a characteristic `camelback' feature in the valence band at the center of the Brillouin Zone ($\Gamma$) due to Spin-Orbit Coupling. This feature is sensitive to the choice of the approximation: it nearly vanishes in GGA, and is over-emphasized in LDA to the extent that the conduction band acquires this feature as well. Many body GW corrections removes this feature and produces a direct gap that has been argued to agree with experiment~\cite{Michiardi2014}. Resolving this controversy is not the main focus of our work, but we note that the inclusion of a GW correction into LDA that ``straightens'' the camelback feature also leads to a substantial reduction of the gap magnitude~\cite{Yazyev2012,Aguilera2013,Michiardi2014}, which is overestimated using LDA by over 50\%, see Table~\ref{Bi2Se3_bulk_LDAGGA}). In contrast, the gap we find using DFT+D2 and similar methods is much closer to the experimental value, and therefore we expect that the features of the band structure we find remain robust to further many-body corrections.

For Bi$_2$Te$_3$, there exists an analogous `camelback' feature in the valence band at $\Gamma$, and  the value of the indirect gap from the DFT+D2 method is close to experiment~\cite{Chen2009}. We find the Valence Band Maximum to occur along the path $L\rightarrow A$ and the Conduction Band Minimum to occur along $A\rightarrow \Gamma$, as seen in figure~\ref{fig:Bulk_Bi2Te3}. While the indirect gap that we find agrees with optical measurements~\cite{Austin1958}, other studies~\cite{Cheng2014,Youn2001,Kim2005} have found the band gap to not occur along lines in reciprocal space which join high-symmetry points in the Brillouin Zone, whereas still others~\cite{Yazyev2012} have found the gap to occur along lines joining high-symmetry points. In the present study, we probed regions of the reciprocal space away from lines joining high-symmetry points, but did not find the band gap to occur close to the points reported in a previous study~\cite{Cheng2014}.

Note also that the band structure exhibits substantial energy dispersion along the $k_z$ direction ($M$-$L$ and $A$-$\Gamma$ in figure~\ref{fig:Bulk}(c) and figure~\ref{fig:Bulk_Bi2Te3}). Therefore, in our opinion,  k-point mesh with only one point along $z$ used in some previous DFT-vdW calculations~\cite{Luo2012,YZhang_vdW:2015} is insufficient for an accurate description of these materials.

\begin{figure}
    \centering
    \includegraphics[scale=0.6]{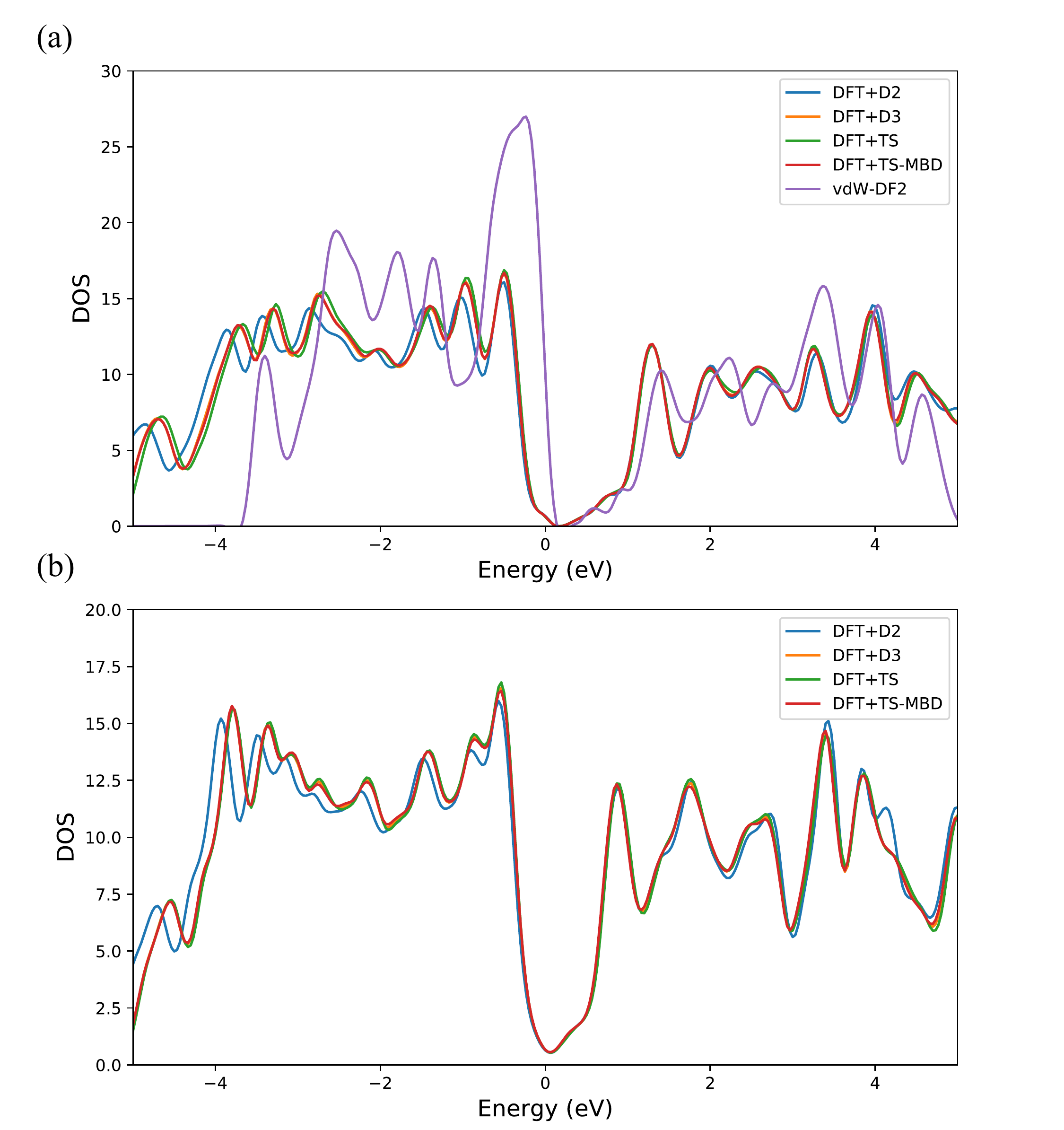}
    \caption{Density of states of bulk Bi$_2$Se$_3$ and bulk Bi$_{2}$Te$_{3}$ using different vdW methods. (a) Bi$_2$Se$_3$ and (b) Bi$_2$Te$_3$.}\label{fig:DOS}
\end{figure}

Our conclusions about the most appropriate methods are further supported by analyzing the Density of States (DOS), shown in figure~\ref{fig:DOS}(a) and figure~\ref{fig:DOS}(b) for Bi$_2$Se$_3$ and Bi$_2$Te$_3$ respectively. The obvious observation is that all semi-empirical methods produce DOS curves which overlap to large accuracy, while the DOS for vdW-DF2 Langreth-Lundqvist method (shown only for Bi$_2$Se$_3$) deviates markedly from the rest, with the valence band approaching the chemical potential, which results in a smaller gap value as listed in Table~\ref{Bi2Se3_bulk_LDAGGA}.

Therefore all the results on the bulk structural and electronic properties show that the semi-empirical methods perform well in reproducing electronic properties of layered topological insulators. In contrast the vdW functionals seem to yield a poor agreement with experiment. 
Recalling that the quintuple layers effectively act as closed shell units, we suggest that the empirical corrections capture the main physics of the vdW interaction between different quintuple layers in Bi$_2$Se$_3$ and Bi$_2$Te$_3$. While all four methods DFT+D2, DFT+D3, DFT+TS and DFT+TS-MBD yield results close to the experiment, we found that DFT+D2 is the most accurate, followed closely by DFT-TS-MBD. The important feature of both methods is that they account only for the long-distance $r^{-6}$ tail of the vdW interactions. This suggests that the vdW correction is indeed due to well-separated charge distributions in different quintuple layers. We choose DFT+D2 for further analysis of the topological surface states.


\section{Slab calculations and the surface states}
We mostly show below the results for the topological surface states obtained using DFT+D2. At first sight, the results are very close to those obtained using DFT+D3, DFT+TS and DFT+TS-MBD. The reason for the closeness, despite DFT+D2 being superior in determination of the structural properties, is that, as discussed below, relaxing the structure of the QLs closest to the surface does not yield appreciable atomic displacement. We expect this to be different for interface calculations, but postpone this analysis to a separate publication. At the same time, we show that, on closer inspection, once again, the DFT+D2 method performs better than its counterparts.

To determine the structure of the surface states we ran calculations for 5-7 QL thick ($\sim$ 50-70 {\AA}) slabs of Bi$_2$Se$_3$ and Bi$_2$Te$_3$, with a vacuum buffer of 100 \AA. 
We found that 5 QL was the minimal slab thickness for which the surface states at opposite surfaces do not hybridize appreciably, so that the Dirac spectrum was not gapped at our energy resolution.

Irrespective of the number of QLs in the slab we consistently found that we needed to include an amount of vacuum that is approximately equal to twice the slab thickness in order to avoid electrostatic interaction of the slab with its own periodic images. Such an interaction generates a gap in the spectrum of the topological surface states, and yield a weak splitting of the top and bottom surface states. The Supplementary Material~\cite{Shirali2020} shows the results for different vacuum thicknesses. While we find that it is possible to fine tune a small vacuum thickness to minimize hybridization of the topological states at the slab interfaces across the vacuum, consistent results are obtained for the vacuum buffer of order twice the slab width. This is in contrast with other work that used $\sim$10$-$20 {\AA} to obtain the surface state dispersion~\cite{Reid2020,Betancourt_2016,Kato2015,Govaerts2014}. We believe this hybridization across vacuum is enhanced by the long tails of the vdW interaction between closed shell QL layers, and checked that the gap at the Dirac point and the splitting of the bands is not affected by the inclusion of dipole corrections at surfaces.

In all our calculations the outermost QLs of the slab (five atomic layers closest to each surface) are allowed to relax, while the atoms in the remaining internal layers (``bulk'') are kept fixed at the optimized bulk structure. We find that the atomic displacements  in the outer layers 
are of order m\AA, not causing appreciable reconstruction, and having negligible effects on the electronic structure. This is consistent with a picture where each QL acts as a closed shell, so that the surface termination does not change bond lengths.


\begin{figure}
    \centering
    \includegraphics[scale=0.9]{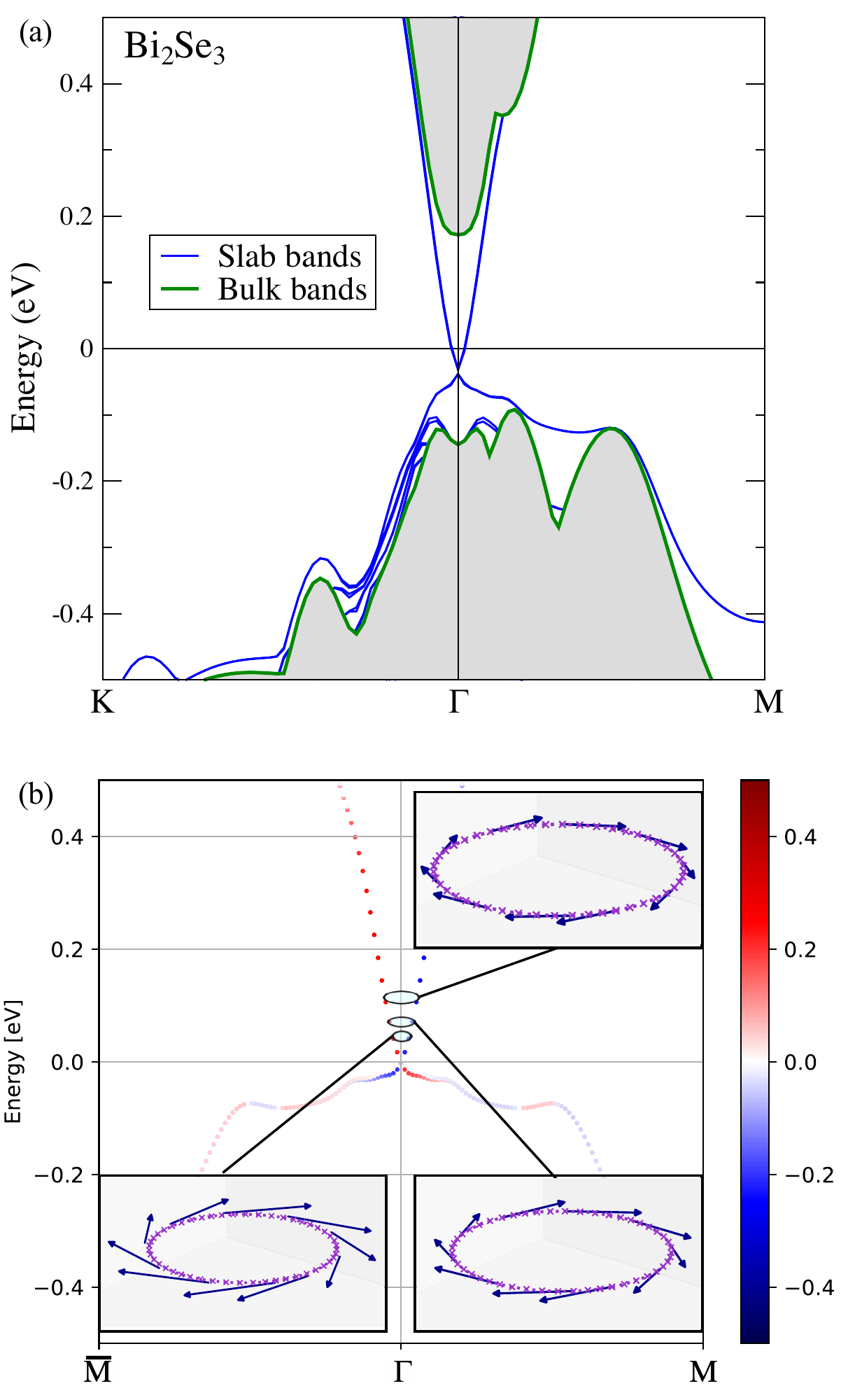}
    \caption{Topological surface states in 5QL thick Bi$_2$Se$_3$ slab with 100 {\AA} vacuum using DFT+D2 van der Waals corrections. (a) Electronic band structure with bulk bands shaded. (b) $y$ component of the spin for the state at the upper surface of the slab along $\bar{M}-\Gamma-M$. Inset: constant energy contours along with the spin textures at those energies, at the upper surface of the slab, for $E = 0.05$ eV, $E = 0.08$ eV, and $E = 0.125$ eV.}\label{fig:surface_state_Bi2Se3}
\end{figure}

\begin{figure}
    \centering
    \includegraphics[scale=0.9]{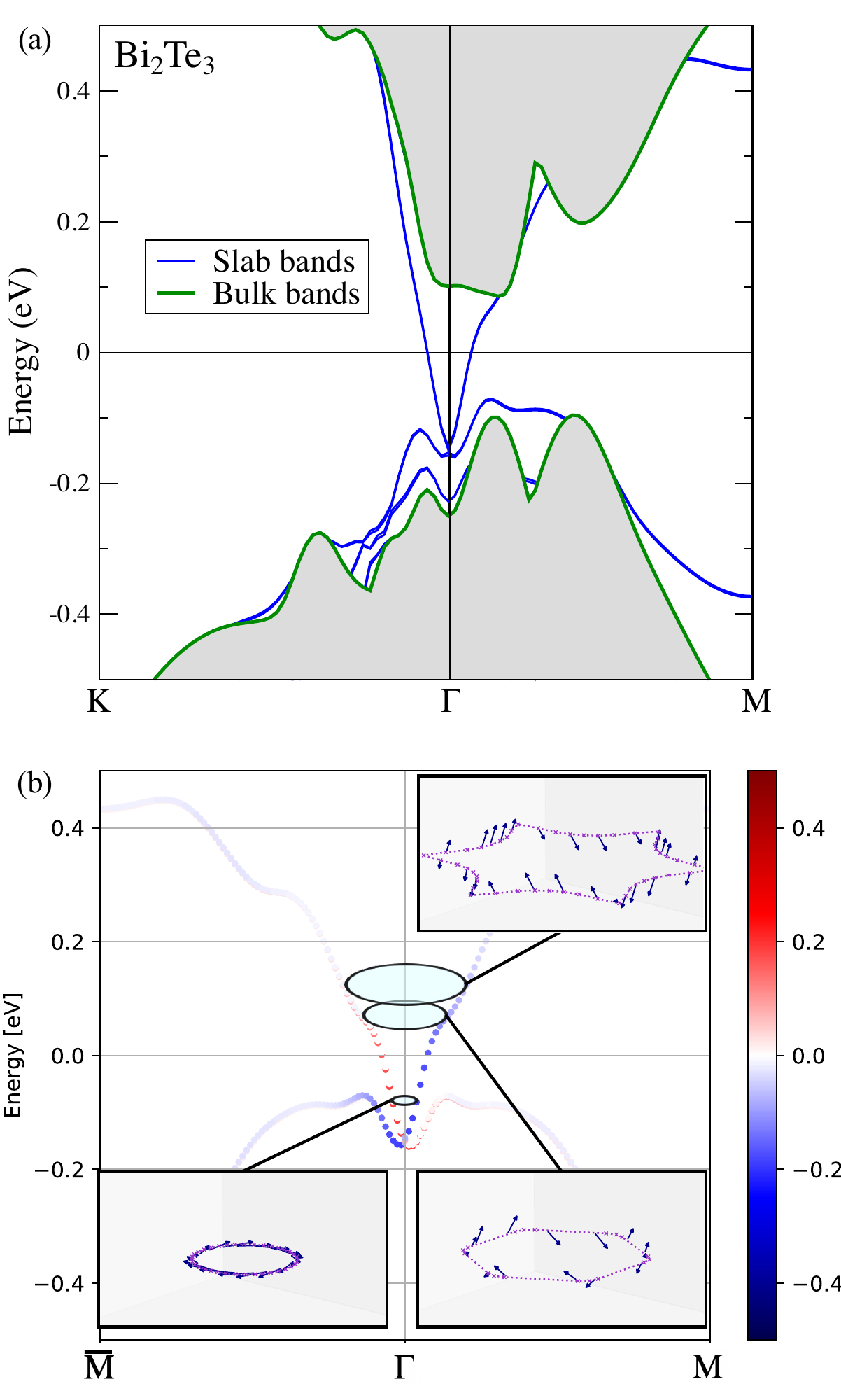}
    \caption{
    Topological surface states in 5QL thick Bi$_2$Te$_3$ slab with 100 {\AA} vacuum using DFT+D2 van der Waals corrections. (a) Electronic band structure with bulk bands shaded. (b) $y$ component of the spin for the state at the upper surface of the slab along $\bar{M}-\Gamma-M$. Inset: constant energy contours along with the spin textures at those energies, at the upper surface of the slab, for $E = -0.12$ eV, $E = 0.03$ eV and $E = 0.07$ eV. The 
    hexagonal warping  increases 
    with increasing energy.}\label{fig:surface_state_Bi2Te3}
\end{figure}

\begin{figure}
    \centering
    \includegraphics[scale=0.9]{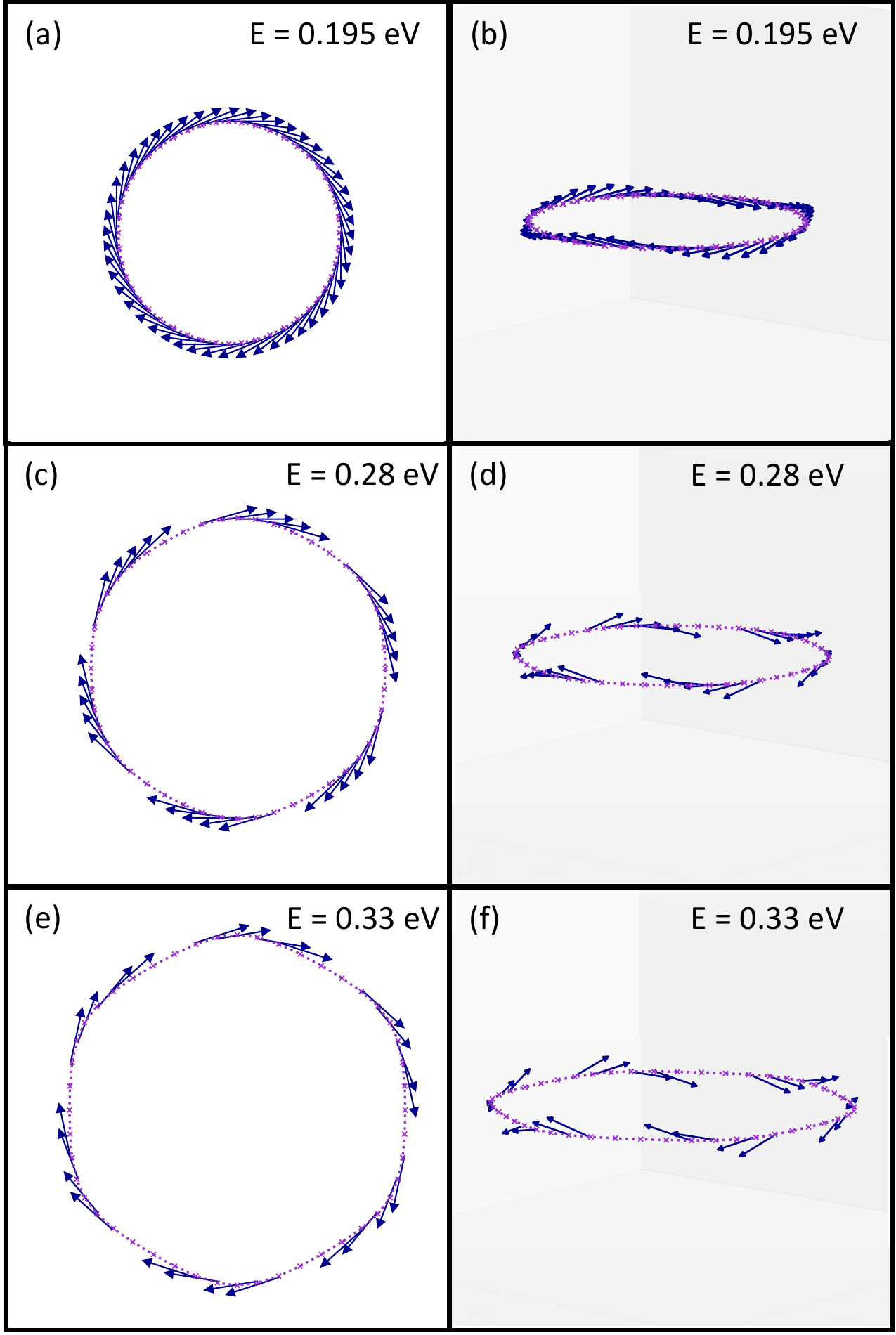}
    \caption{Spin textures and constant energy contours of the topological surface states at energies near and above the conduction band minimum, at the upper surface for a 5 QL thick slab of Bi$_2$Se$_3$ with 100 {\AA} vacuum, calculated using DFT+D2 van der Waals corrections. Spin textures in the $k_x$-$k_y$ plane are shown at energies (a) $E = 0.195$ eV, (c) $E = 0.28$ eV, and (e) $E = 0.33$ eV. Out-of-plane components of the spin textures are shown at energies (b) $E = 0.195$ eV, (d) $E = 0.28$ eV, and (f) $E = 0.33$ eV. The constant energy contours begin to display hexagonal warping at energies of around $0.28$ eV and above.}
    \label{Bi2Se3_additional_spin_textures}
\end{figure}

The slab band structures for Bi$_2$Se$_3$ (Bi$_2$Te$_3$) are shown in figure~\ref{fig:surface_state_Bi2Se3}(a) (figure~\ref{fig:surface_state_Bi2Te3}(a)).
They show quasi-linearly dispersing Dirac-like states in the bulk gap arising from the surface states. For Bi$_2$Se$_3$, the Dirac point lies $0.21$ eV below the conduction band minimum, in agreement with experiments~\cite{Xia2009,Analytis2010,Hsieh2009,Chen2010}, and $0.06$ eV above the valence band maximum. This is in qualitative agreement with experiment~\cite{Zhang2009}, although self-doping effects likely affect the location of the Dirac point. For Bi$_2$Te$_3$, the surface band rises above the Dirac point energy as we move along the path $\Gamma \rightarrow M$, a feature not seen in Bi$_2$Se$_3$. The Dirac point lies $0.05$ eV below the valence band maximum (VBM), and $0.23$ eV below the conduction band minimum. The latter value is lower than $0.26$ eV obtained in an LDA calculation, and closer to the GW value of $\sim 0.22$ eV~\cite{Forster2016}. All these values are less than $0.29$ eV obtained from optical measurements~\cite{Chen2009}. The energy difference between the Dirac point and VBM was theoretically found to be $0.10$ eV in a previous study~\cite{Yazyev2012} and experimentally determined to be $0.13$ eV~\cite{Chen2009}.


To compare different semi-empirical vdW methods we evaluate the velocities of the surface Dirac quasiparticles, $v$. Since the dispersion is quasi-linear, we choose to compute it
along the path $\Gamma-M$ close to the k-point $\vec{k}=0.015\vec{b}_1$, and show the computed and experimental values in Table~\ref{Dirac_velocities}.
For Bi$_2$Se$_3$, the value $v=5.1 \times 10^{5}$ m/s obtained using the DFT+D2 method, is much closer to the experimental range of $5.0-5.5\times 10^{5}$ m/s ~\cite{Devidas2017,Pertsova2016,Kuroda2010,Chang2015,Cao2013} than the values obtained using other methods (DFT+D3, DFT+TS, DFT+TS-MBD). Our value is also much closer to experiment than those reported in previous ab initio calculations~\cite{Yazyev2010,Crowley:2015}. For Bi$_2$Te$_3$, the velocities calculated using different vdW implementations are nearly identical, and the velocity obtained using the DFT+D2 method is close to the values of $3.87 \times 10^{5}$ m/s and $4.0 \times 10^{5}$ m/s reported in experiment~\cite{Chen2009,Qu2010}.

\begin{table}
\caption{\label{Dirac_velocities} Calculated Dirac velocities of the surface states in slab calculations of Bi$_2$Se$_3$ and Bi$_2$Te$_3$ using different forms of van der Waals corrections.}
\begin{indented}
\item[]\begin{tabular}{@{}|l|l|l|}
\br
   & Bi$_2$Se$_3$ & Bi$_2$Te$_3$\\
 \hline
  & $v$ (m/s) & $v$ (m/s)\\
 \hline
 Expt. & $5.0\times 10^{5}$~\cite{Xia2009} & $4.0\times 10^{5}$~\cite{Qu2010}\\
 \hline
 GGA & $4.80 \times 10^{5}$ & $3.83\times 10^{5}$\\ 
 \hline
 DFT+D2 & $5.10\times 10^{5}$ & $4.04\times 10^{5}$\\ 
 \hline
 DFT+D3 & $4.07\times 10^{5}$ & $3.96\times 10^{5}$\\ 
 \hline
 DFT+TS & $4.00\times 10^{5}$ & $4.19\times 10^{5}$ \\
 \hline
 DFT+TS-MBD & $3.77\times 10^{5}$ & $4.42\times 10^{5}$\\ 
 \br
 \end{tabular}
 \end{indented}
 \end{table}

Analyzing the spin structure of the topological states, we find, as expected, that the surface states are nearly perfectly helical close to the $\Gamma$ point, as shown in  figure~\ref{fig:surface_state_Bi2Se3}(b) (figure~\ref{fig:surface_state_Bi2Te3}(b)) for Bi$_2$Se$_3$ (Bi$_2$Te$_3$). The spins are normal to the direction of the momenta, and wind around the Dirac point. At higher energies, however, the difference between the two materials becomes evident.  The insets in figure~\ref{fig:surface_state_Bi2Se3}(b) and figure~\ref{fig:surface_state_Bi2Te3}(b) show the {\it computed} spin orientations along different constant energy surfaces. For Bi$_2$Se$_3$, we find that at energies close to the Dirac point, the spins are helical, and there is no discernible evidence of hexagonal warping of the constant energy contours (for example, at $E=0.125$ eV), in agreement with experiments~\cite{Nomura2014,Kuroda2010}. Only at energies close to $0.3$ eV (where the Fermi level often lies in real Bi$_2$Se$_3$ films~\cite{Yang2019,Zhang2010,Analytis2010,Hsieh2009}) and higher, when the energy is in the conduction band, the constant energy contour begins to take on a hexagonal shape, and the spins begin to develop out-of-plane components, as shown in figure~\ref{Bi2Se3_additional_spin_textures}. In contrast, for Bi$_2$Te$_3$, 
the constant energy contour changes shape with increasing energy at much lower energies, changing from a circle ($E = -0.12$ eV), to a hexagon ($E = 0.03$ eV), to a star ($E = 0.07$ eV). At the same time the spins acquire a substantial out-of-plane component. It is clear therefore that the hexagonal warping and non-helical behavior of the topological surface states appear together, but that happens at different energies in the two materials.

\begin{figure}
    \centering
    \includegraphics[scale=1.1]{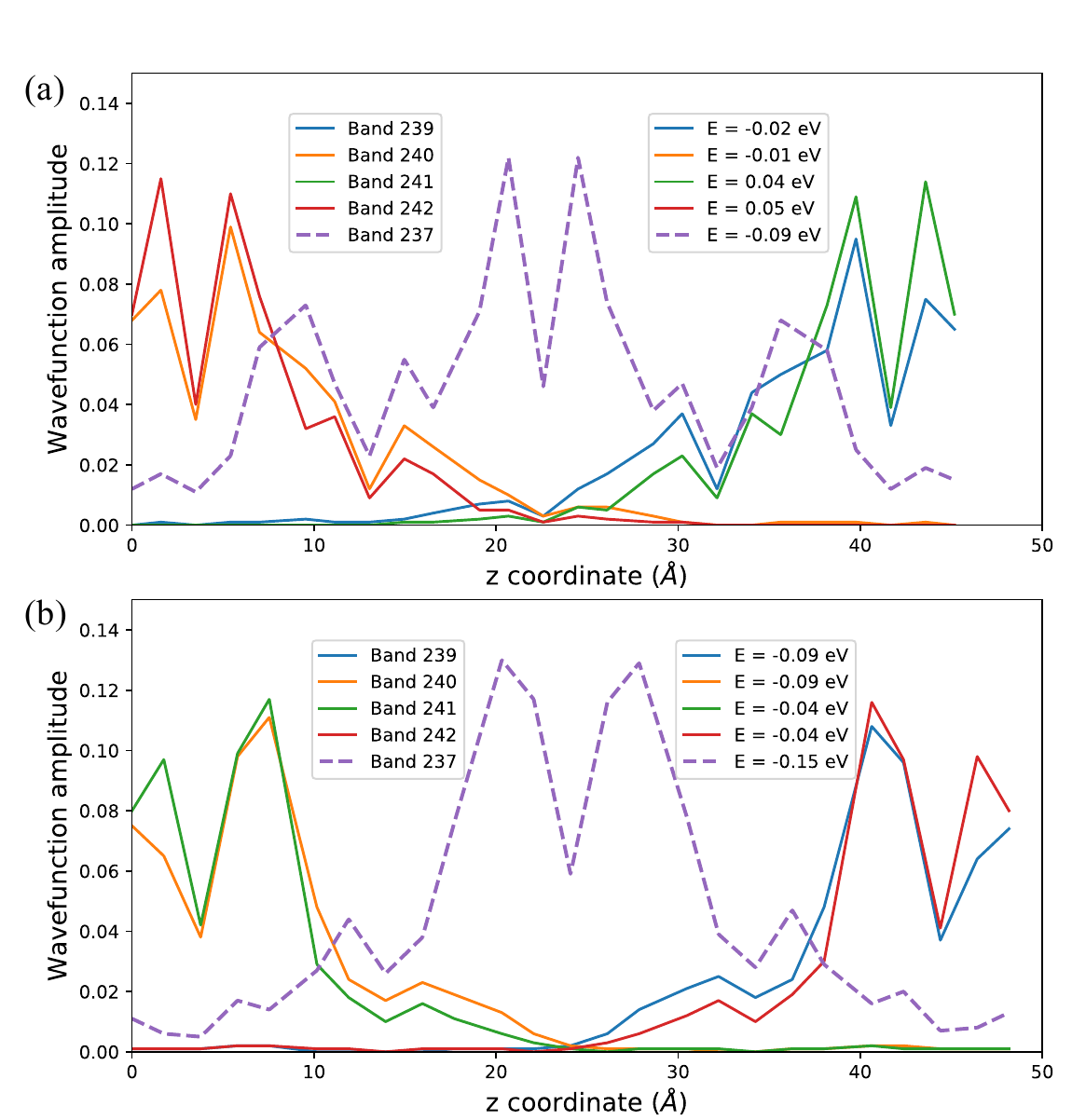}
    \caption{Spatial profile of surface (solid lines) and bulk (dashed lines) states  (smooth lines) and a state in the bulk bands (dashed lines) at $\bm{k} = 0.0127 \; \bm{b}_1$, for 5QL thick slabs with 100 {\AA} vacuum.  $\bm{b}_1$ is the in-plane reciprocal lattice vector for the hexagonal unit cell choice.  Band 239/240 (241/242) belong to the upper (lower) half of the Dirac cone, while band 237 is near the top of the valence band.  (a) Bi$_2$Se$_3$; (b) Bi$_2$Te$_3$.}
    \label{fig:surface_state_amplitude}
\end{figure}

Figure~\ref{fig:surface_state_amplitude} shows the spatial profiles of the probability amplitude for the surface states in comparison with the band states. While all states oscillate on atomic scale, the surface states are localized within approximately a single QL near the interfaces, while the amplitude of the band states is peaked in the central, bulk-like, QLs.

\section{Discussion} We performed a comprehensive high precision ab-initio analysis of the prototypical topological insulators Bi$_2$Se$_3$ and Bi$_2$Te$_3$, aiming to treat structural and electronic properties on equal footing. We used different methods of accounting for the  van der Waals interactions within the framework of the density functional theory. In agreement with the expectation that vdW interactions play an important role in binding closed shell quintuple layers in layered topological insulators, we found that, in general, their inclusion  strongly influences the structural properties, if the structure is allowed to relax and not fixed to the experimental values.  Lattice constants, unit cell volume, and the distance between nearest quintuple layers are all sensitive to the inclusion and choice of the vdW interaction.

Standard exchange-correlation functionals such as LDA and GGA, are inadequate for structural relaxations, leading to structural and electronic properties which deviate significantly from the experimental values.

We compared two different classes of methods that include vdW interactions in first principles calculations: those which include semi-empirical corrections to the total energy and forces, and those which modify the exchange-correlation functionals to include a vdW term. Our results strongly indicate superiority of the semi-empirical methods, which yield lattice parameters, energy gap, bulk density of states, and the Dirac velocity of the topological surface states in much better agreement with the experimental values.
By comparing the errors in the lattice constants vs the inter-QL spacing, we conclude that the long-range energy correction due to interaction between QLs, is the dominant effect of van der Waals forces in layered TIs.
Of the semi-empirical methods, the two which account solely for the long-range tail  $r^{-6}$ of the vdW interaction (DFT+D2, and DFT+TS-MBD) give much better structural parameters than the methods that attempt to augment it further by including also a shorter range $r^{-8}$ correction (DFT+D3). To us this indicates that the dominant vdW interaction is between the quintuple layers, which act as closed shells.

Extending the comparison to include the electronic properties, we found that the DFT+D2 method in particular gives both very accurate electronic parameters, and yields dispersion and spin structure of the topological surface state that closely match experimental results. Overall, we observe an order of magnitude improvement in the lattice parameters in comparison to when van der Waals corrections are not included; we also obtain accurate values of the Dirac velocities, which indirectly implies that the DFT+D2 method is able to reproduce fine aspects of the electronic structure on scales smaller than the band gap. 
Therefore our result identifies the leading correction to the commonly used DFT functionals for ab initio calculations of surfaces and interfaces involving topological insulators of the Bi$_2$X$_3$ family.

We found that the vdW-DF2 functional yields structural and electronic parameters (cell volume and density of states) for the bulk which deviate significantly from experiments. Our interpretation of this result is that in this class of systems, modification of the exchange-correlation functional is not needed and only the vdW tails are important. The SCAN functional predicted the bulk 3D TIs to be metals.

We believe that the semi-empirical methods, such as DFT+D2 and TS succeed because the materials we study have two characteristic qualities: closed shell QLs, where all bonds are saturated within the QL, and the long-range $r^{-6}$ correction is the most important, and the 3D band structure, where the vdW forces are a correction but not the principal source of the dispersion normal to the layers. The former is backed up by our observation that DFT+D3, which includes an $r^{-8}$ correction in addition to the $r^{-6}$ correction, does not perform as well as DFT+D2. Thus, for materials which possess both the attributes of closed shell layers and 3D interlayer dispersions, we expect such methods to yield accurate results. They are unlikely to perform as well in layered materials which do not possess both of these features, such as the transition metal dichalcogenides. 

Methodologically, we found that inclusion of the long range van der Waals interactions requires a vacuum buffer of roughly twice the slab thickness in the system to accurately describe surface states. This suggests that in the calculations of interfaces and heterostructures one needs to pay special attention to the decay range of the interactions with the material thickness. Thus, for calculations involving surfaces, an appropriate amount of vacuum must be included in the setup so as to not obtain erroneous results due to electrostatic interaction of the surfaces across the vacuum. We have also demonstrated here that the interlayer distance is tied to the strength of the vdW interactions in the system. Thus, obtaining inaccurate interlayer distances in a calculation is a possible indication that crucial vdW corrections are missing; if a heterostructure is constructed with those structural parameters, the values of stress and strain obtained at the interface are likely to be inaccurate.

While previous studies have included select vdW interactions in first principles calculations of specific aspects of the properties of topological insulators, to our knowledge there have not been many systematic investigations of the different ways of accounting for the  van der Waals corrections. Determining the optimal vdW method is necessary for any calculation of TIs where the stress and strain fields must be calculated self-consistently. This would be especially important for ab-initio calculations of heterostructures based on topological insulators, where accounting for the  stress and strain fields at the interface is crucial for predicting and analyzing possible surface reconstruction and concomitant changes in both dispersion and the spin-momentum locking properties of the topological states. Thus, the simultaneous optimization of the structural and electronic properties is essential for determining the properties of topological insulators.

To reiterate our main point, when performing calculations on proposed but not yet synthesized or not well characterized interfaces, a systematic approach that avoids assumptions about structural or electronic parameters is necessary. Making such assumptions is an approach that works a posteriori for the surface, but one that we explicitly deemed unsuitable for interface calculations, because of the loss of predictive power for systems where we do not have a priori knowledge of the lattice parameters. Such assumptions about structural or electronic parameters might result in obtaining a solution which is not the ground state. Thus, including the DFT+D2 vdW correction would be appropriate and important when investigating heterostructures of the 3D TIs studied here with other materials. If including DFT+D2 corrections in the other material leads to correctly reproduced bulk structural and electronic properties, it is straightforward to include this correction for the entire system. On the other hand, for systems involving interfaces between 3D TIs and materials in which vdW corrections are not needed, a suitable transition region could be constructed at the interface, over which the strength of the vdW correction goes to zero. When studying the topological interface states in such heterostructures, properties such as the Dirac velocity and the spin structure of surface states will be affected by the inclusion and type of the vdW corrections used, as we have shown here for surfaces.

In summary, our results establish a pathway for computationally efficient and reliable consistent determination of the surface and interface properties of topological materials, and for ab initio analysis of prototype topological devices including stress, strain, and, in principle, symmetry breaking effects due to atomic reconstruction. 

\section*{Acknowledgements}
This research was supported by NSF via Grant No. DMR-1410741 (K. S. and I. V.) and by the U.S. Department of Energy under EPSCoR Grant No. DE-SC0012432 with additional support from the Louisiana Board of Regents (W.~A.~S.). Portions of this research were conducted with high performance computing resources provided by Louisiana State University (\texttt{http://www.hpc.lsu.edu}).

\providecommand{\newblock}{}

\end{document}